\documentclass [aip,jmp, amsmath,amssymb,reprint]{revtex4-1}

\usepackage{amssymb}
\usepackage{amsbsy}
\usepackage{amsmath}
\usepackage{amsfonts}
\usepackage{mathrsfs}
\usepackage{latexsym}
\usepackage[english]{babel}
\usepackage{url}
\usepackage[]{graphicx}
\usepackage{epstopdf}
\usepackage{subfigure}
\usepackage{hyperref}

\begin{document}

\title{Ferromagnetic resonance and magnetoresistive measurements evidencing magnetic vortex crystal in nickel thin film with patterned antidot array}

\author{I.\ R.\ B.\ Ribeiro}
\affiliation{ Departamento de F\'{i}sica, Universidade Federal de Vi\c cosa, Vi\c cosa, 36570-900, Minas Gerais, Brazil }
\affiliation{ Instituto Federal do Esp\'{i}rito Santo, Alegre, 36570-900, Esp\'{i}rito Santo, 29520-000, Brazil }

\author{J.\ F.\ Felix}
\affiliation{ Departamento de F\'{i}sica, Universidade Federal de Vi\c cosa, Vi\c cosa, 36570-900, Minas Gerais, Brazil }
\affiliation{ Instituto de F\'{i}sica, N\'{u}cleo de  F\'{i}sica Aplicada, Universidade de Bras\'{i}lia-UnB,  Bras\'{i}lia, 70910-900, Distrito Federal, Brazil }

\author{L.\ C.\ Figueiredo} 
\affiliation{ Instituto de F\'{i}sica, N\'{u}cleo de  F\'{i}sica Aplicada, Universidade de Bras\'{i}lia-UnB,  Bras\'{i}lia, 70910-900, Distrito Federal, Brazil }

\author{P.\ C.\ de Morais} 
\affiliation{ Instituto de F\'{i}sica, N\'{u}cleo de  F\'{i}sica Aplicada, Universidade de Bras\'{i}lia-UnB,  Bras\'{i}lia, 70910-900, Distrito Federal, Brazil }
\affiliation{ Huazhong University of Science and Technology, School of Automation, 430074, Wuhan, China}

\author{S.\ O.\ Ferreira}
\affiliation{ Departamento de F\'{i}sica, Universidade Federal de Vi\c cosa, Vi\c cosa, 36570-900, Minas Gerais, Brazil }

\author{W.\ A.\ Moura-Melo}
\affiliation{ Departamento de F\'{i}sica, Universidade Federal de Vi\c cosa, Vi\c cosa, 36570-900, Minas Gerais, Brazil }

\author{A.\ R.\ Pereira}
\affiliation{ Departamento de F\'{i}sica, Universidade Federal de Vi\c cosa, Vi\c cosa, 36570-900, Minas Gerais, Brazil }

\author{A.\ Quindeau}
\affiliation{ Department of Physics and Francis Bitter Magnet Lab, Massachusetts Institute of Technology, Cambridge, MA 02139, USA }

\author{C.\ I.\ L.\ de Araujo}
\email{dearaujo@ufv.br} 
\affiliation{ Departamento de F\'{i}sica, Universidade Federal de Vi\c cosa, Vi\c cosa, 36570-900, Minas Gerais, Brazil }

\begin{abstract}
Ferromagnetic vortices deliver robust out-of-plane magnetization at extremely small scales. Their handling and creation therefore has high potential to become a necessary ingredient for future data storage technologies in order to keep up with the pace of growing information density demands.  In this study we show that by using one step nanolithography method, we are able to create ferromagnetic vortex lattices in thin nickel films. The necessary control of the magnetic stray field at the domain edges was achieved by actively modifying the ferromagnetic thin film anisotropic properties at nanometer scale. We present experimental evidence using ferromagnetic resonance and magnetoresistance measurements supporting simulations based on the theoretical prediction of the proclaimed vortex structures.\par

\end{abstract}


\maketitle

Anisotropic magnetoresistance (AMR) occurs as a weak phenomenon in any soft ferromagnetic material. The exploitation of this effect enabled the development of the first industrial magnetic sensors. However, due to fundamental limitations owing to the spin-orbit interaction \cite{AMR}, junctions based on AMR did not exceed the 1\%-mark. Later advancements in microstructure design allowed the creation of heterostructures involving multiple layers of magnetic and nonmagnetic conducting materials. As a result, the "giant" magnetoresistance (GMR) was discovered, leading to a huge increase in the magnetic sensor sensitivity \cite{Fert, Grumberg}. Innovative material compositions that introduced antiferromagnetic pinning layers to (exchange-) bias the coercive fields of the ferromagnetic films opened the door for the design of very efficient and reliable magnetic switches\cite{Dieny1}.\par
The necessity to produce hard disc drives with increasing storage density to cope with the advancements of the computer industry and the internet demanded even larger magnetoresistive effects. Following the trend of scaling down computer architecture to the nanometer range, the nonmagnetic conducting films in GMR junctions were replaced by ultra thin insulating oxide films to allow electronic transport only via quantum tunneling. The spin filtering properties of those tunnel magnetoresistance (TMR) junctions hence increased dramatically \cite{moodera1, moodera2}. Benefiting from this very high magnetic sensitivity, devices such as the magnetoresistive random-access memory (MRAM) became feasible. 
Today, new challenges arise as a product of recent industrial developments. Improving the speed and durability of magnetic recording as well as reducing the power consumption becomes important to a growing number of applications.

\begin{figure}
\center
\includegraphics[width=7cm]{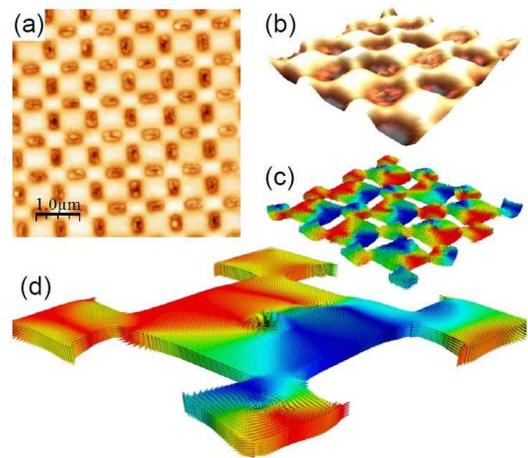}
\caption{\label{fig1} (a) Atomic Force Microscopy (AFM) image of the sample, (b) AFM three-dimensional visualization (c) Micromagnetic simulation of the ground state after magnetization relaxation with randon quirality and polarization. (d) Magnification of the spin configuration inside a single vortex.}
\end{figure}
\begin{figure*}
\center
\includegraphics[width=17cm]{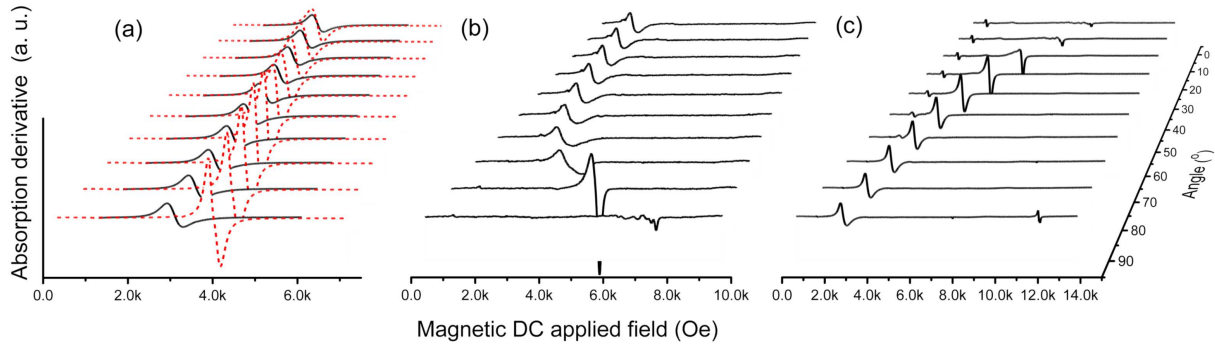}
\caption{\label{fig2} (a) Ferromagnetic resonance measurements data obtained with applied magnetic field in unpatterned $Ni$ thin film in-plane sample (black lines) and out-of-plane (red dots), with rotating angles from (90${\rm ^{o}}$) to (0${\rm ^{o}}$), (b) in-plane measurements performed in patterned sample and (c) out-of-plane measurements performed in patterned sample.}
\end{figure*}
Approaches to potential solutions include the magnetic domain-wall racetrack memory \cite{Parkin} and magnetic dots \cite{dot1,dot2,dot3}. Among the latter, nanodisks appear to be very promising candidates once they allow magnetic vortices as stable states \cite{cowburn,disk1,disk2}, which would enable controllable switching between a couple of holes and thus yield tiny logic binary elements \cite{Rahm}. Large arrays of nanodisks have been proposed to complement thermally assisted MRAM devices in order to improve stability and to avoid magnetostatic traps, as they occur in ordinary rectangular elements \cite{Dieny2,Dieny3} and thus demand relatively strong readout fields.\par
In this work, we corroborate experimentally the theoretical prediction that suitably patterned array of platters, supports a magnetic crystal vortex, due to the anisotropy generated by the stray field on its antidot borders \cite{APL2014}. This system brings advantages in relation to arrays of nanodisks (conventional elements that support a vortex state), once the vortex are electrically connected in the same material, resulting in low power consumption in vortex manipulated by alternate currents \cite{vortexac}. We characterize the as-fabricated samples using ferromagnetic resonance (FMR) and magnetoresistance (MR) measurements and bring the observed behavior in context with theoretical predictions of the vortex configuration and spin dynamics \cite{FMRdisks}.\par
For patterning, a silicon substrate previously covered with a $250 nm$ polymethylmethacrylate (PMMA) film was brought into a RAITH e-LINE Plus system, where the exposure of the antidot design was performed. This procedure was repeated several times to form shapes with area of ${100\, \mu \textrm{m}^{2}}$ (see Fig. \ref{fig1} a and b). 
After the development of the PMMA film, the samples were placed into a Thermionics e-Beam evaporation system where a $30 nm$ nickel ($Ni$) film was deposited on top of a $5 nm$ titanium seed layer. A gold capping layer of $3 nm$ was deposited on top of the sample to prevent $Ni$ oxidation. In this step, we also performed a sample with same thickness but without patterning for comparison. The structures were eventually finalized by a lift-off process in an acetone ultrasonic bath. The magnetization dynamics of the as-fabricated samples have been investigated via FMR carried out in a X-band spectrometer (Bruker EMX PremiumX, equipped with an ER 4102ST resonator). For the MR measurements electric contacts were developed by photolithography defining regions where a $50 nm$ thick gold was deposited. Subsequently, a lift-off process provided two electrodes in the borders of the antidots samples with channels of ${80\, \mu \textrm{m}}$ length, on which a direct current of $10 mA$ was applied.
Micromagnetic simulations performed with the software Object Oriented MicroMagnetic Framework ($OOMMF$) \cite{oommf}, utilizing square mesh of $5 nm$ edge and parameters for $Ni$ as saturation magnetization $4.3$x$10^5 A/m$, exchange constant $9$x$10^{-12} J/m^3$, cubic anisotropy constant $-5.7$x$10^3 J/m^3$ and damping coefficient $0.01$, provided ground state magnetization of vortex crystal configuration in which chirality and polarization appear to be randomly distributed throughout the sample (see Fig. \ref{fig1} c and d). Upon application of external magnetic field in parallel to the sample plane (in-plane configuration), a relatively small magnetic field is sufficient to annihilate the vortex patterns, as demonstrated in previous work \cite{vortexanhilation, APL2014}. Furthermore, the spins seem to relax and align with the external applied magnetic field except at the borders, where the spin-stray field tends to pin the spins parallel to each border \cite{Laura}. Once two adjacent borders are perpendicular to each other, the spin-stray field coupling may render the sample to be magnetically anisotropic. Indeed, in the comparison between experimental FMR performed in uniform $30 nm$ $Ni$ thin film and in patterned sample in function of rotating angle in-plane and out-of-plane, depicted in Fig. \ref{fig2}, is possible to note that in the thin film the peak of resonance field ($H_R$) is isotropic in-plane, see Fig. \ref{fig2} a, while presents a slight change in out-of-plane configuration due to the shape anisotropy. However, in the measurements performed in patterned sample in-plane configuration, presented in Fig. \ref{fig2} b, the value for $H_R$ isotropic peak is lower, around $1.1 \rm{kOe}$ (at $0^o$), which shifts towards higher values as the applied magnetic field in-plane angle progresses towards $90^o$, thus accounting for the pinning of spins at the borders of the platters. This behavior is similar to the commonly observed one in anisotropic magnetic systems \cite{nanowires}.
When the FMR measurements were performed starting with the external magnetic field applied out-of-plane, besides the FMR peaks at $1.1 \rm{kOe}$, a huge peak observed at $2.4 \rm{kOe}$ (Fig. \ref{fig2} c) can be attributed to the resonance field of vortex core polarizations. This, therefore, tend to align and remain very stable along the applied field axis, similarly to what happens in nanodisk arrays \cite{polarizationandfield}. At magnetic field strengths of around $11.6 \rm{kOe}$, in the out-of-plane saturation magnetization regime, more FMR peaks are observed, suggesting formation of magnon excitations, as shown in Fig. \ref{fig2} c and in more detail in Fig. \ref{fig4} c. A similar behavior was observed in perpendicular FMR spectrum analysis on micrometer-sized disks, recently reported by Castel et al. \cite{FMRdisks}. It was shown that in the saturated regime, one main resonance line along with several peaks decreasing in amplitude on the low-field side is observed in the disk array. These multiple resonance peaks have been attributed to standing in-plane spin-wave modes. \par
\begin{figure}[h]
\centering
\includegraphics[width=7.5cm]{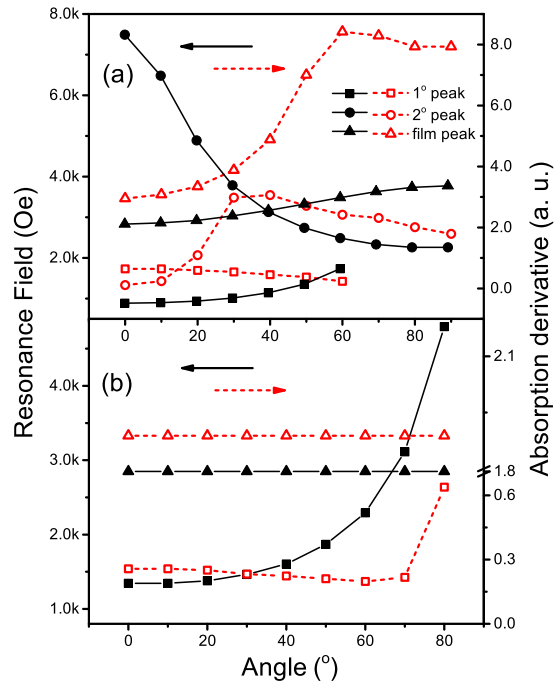}
\caption{\label{fig3} (a) Resonance field and peaks intensity performed in out-of-plane configuration as a function of angle in patterned sample and $Ni$ thin film and (b) Resonance field and peaks intensity performed in-plane configuration. }
\end{figure}
\begin{figure}
\center
\includegraphics[width=8.5cm]{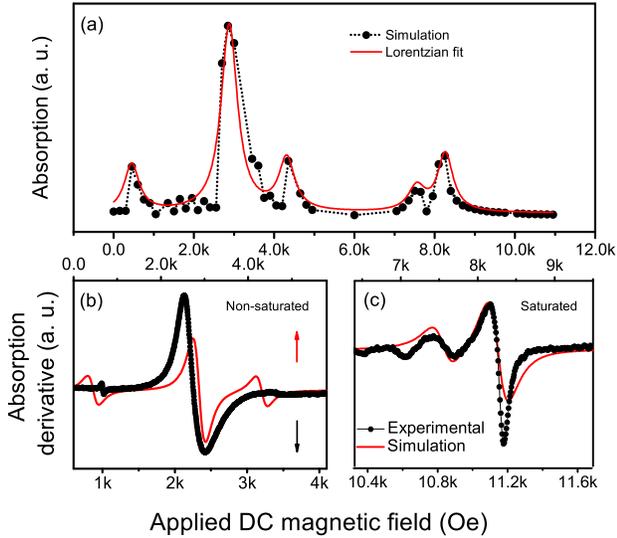}
\caption{\label{fig4} (a) Data obtained from dynamics micromagnetic simulation and lorentzian fit, (b) derivative absorption obtained from the fit in (a) and comparison in non-saturated regime experimental FMR data  and (c) derivative absorption obtained from simulation and comparison with the saturated regime.}
\end{figure}

In the present context, the externally applied magnetic field is divided into two components, namely in-plane and out-of-plane contributions. As the effective out-of-plane component decreases with tilting angle, $H_R$ increases from $2.4 \rm{kOe}$ to $6.5 \rm{kOe}$, as a result of fixed core orientation along out-of-plane axis. On the other hand, the in-plane component  is responsible for increase in magnetization of chiral spins aligned along the external field axis, which causes vortex quenching bellow 15${\rm ^{o}}$. For better observation, the behavior of $H_R$ (black) and peak amplitudes (red dash) in function of rotating angle in both in-plane and out-of-plane configuration, for patterned sample and $Ni$ thin film are summarized in Fig. \ref{fig3}, where squares and circles represent the first and second peak, respectively, observed in the patterned sample, while triangles represent the peak in $Ni$ thin film. 
From this data is possible to follow the similar behavior of thin film $H_R$ peak and first peak in the patterned sample that is suppressed by the vortex peak after 60 ${\rm ^{o}}$. In our experiment, the $AC$ microwave magnetic field is perpendicular to the $DC$ sweeping field. As the absorption peak intensity is related to the configuration of microwave incidence in the thin film samples, due to the magnetic material interaction, the values will be always lower in-plane than in out-of-plane configuration, as can be notice in a comparison of values in figures \ref{fig3} a and \ref{fig3} b. The intensity increase in thin film peak with saturation after 60 ${\rm ^{o}}$, represented by red triangles in Fig. \ref{fig3} a, can be related to the alignment of sample plane with the $AC$ microwave magnetic field and the second peak intensity represented by red circles behaves like the first peak (red squares), after vortex quenching. In Fig. \ref{fig3} b the isotropy expected for peak intensity and $H_R$ in the thin film in-plane configuration was confirmed, while the spin pinning in the borders of antidots gives rise to anisotropy in the patterned sample. \par
\begin{figure}
\centering
\includegraphics[width=8cm]{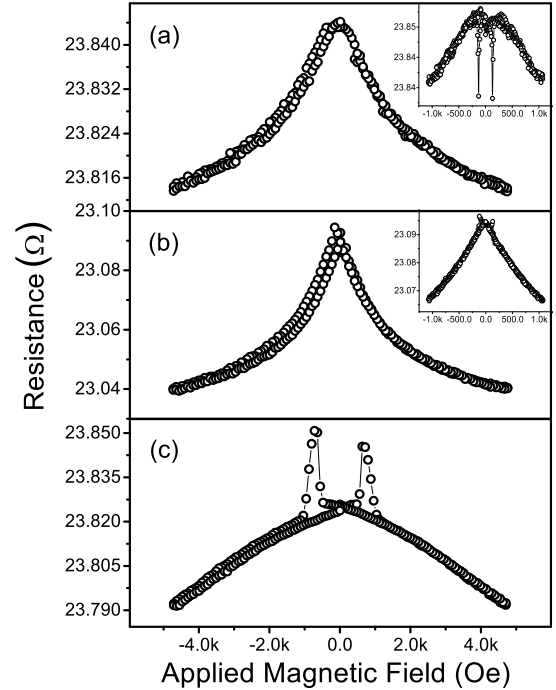}
\caption{\label{fig5}(a) Magnetoresistance measurements performed in the antidot sample in configuration of applied DC electric current (a) Longitudinal, (b) Transversal and (c) perpendicular to the applied magnetic field. In the insets are presented zoom of the peaks with the anisotropic magnetoresistance signal expected for ferromagnetic thin films. }
\end{figure}

\begin{figure}
\centering
\includegraphics[width=7cm]{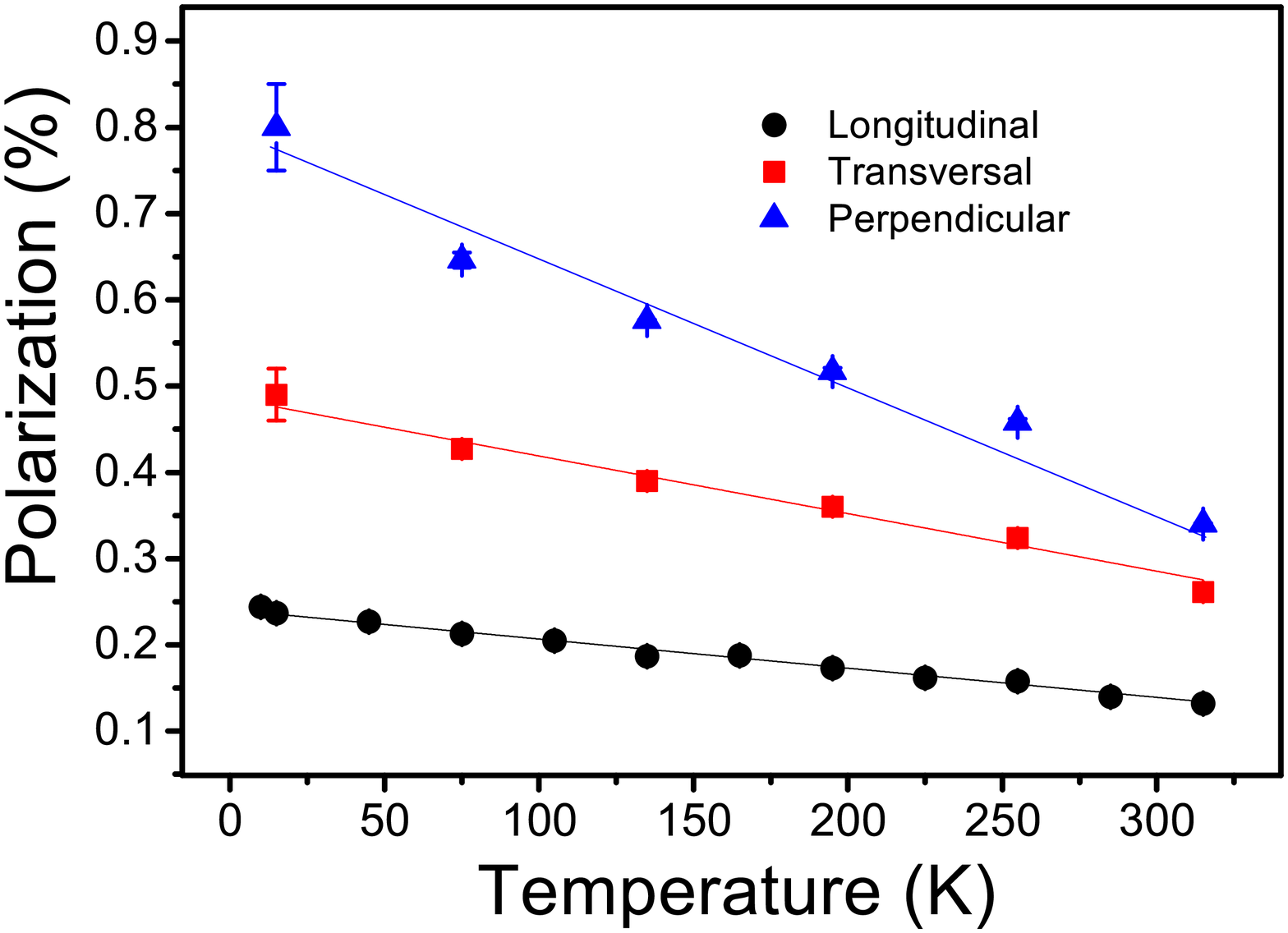}
\caption{\label{fig6} Long range spin polarization measured by local magnetoresistance as a function of temperature. }
\end{figure}

 
Our observations were corroborated by theoretical dynamics investigations performed by micromagnetic simulation in $Mumax^3$ GPU based code\cite{mumax}. With same $Ni$ parameters described before and following procedure utilized in references \cite{mic1,mic2}, we have applied an alternating field, which is bases upon the following equation:\begin{equation}\vec B^{ac}_y=(1-e^{\lambda t})\vec B_{ac,0}cos(\omega t)\end{equation}
, with same intensity $\vec B_{ac,0}=10 Oe$ and $\lambda \sim f=\omega / (2\pi)=9.5 GHz$, utilized in experiment. For each step of $DC$ $Bz$, 1000 interactions during $5ns$ were performed. After discard the 600 first interactions and apply Fourier transform in the stable $m_y$ signal, we have obtained the absorption data presented in Fig. \ref{fig4} a. The derivative of the data lorentzian fit gives the FMR spectra, which is compared with non-saturated regime experimental data in Fig. \ref{fig4} b and saturated regime in Fig. \ref{fig4} c. The main peaks observed experimentally were also present in the simulations, despite of the difference between the experimental and theoretical $H_R$ and peak intensity, which can be attributed to the zero temperature in the simulation and defects in sample borders and antidots that can act as magnetostatic traps for spin pinning.  
Additional experimental evidence of the topological crystal vortex based on MR measurements is presented in Fig. \ref{fig5}. The MR measurements were carried out with the applied magnetic field in the longitudinal and transverse configuration (in-plane) and in the perpendicular configuration (out-of-plane) at $300 K$.
As it can be seen on the magnified areas in Fig. \ref{fig5}, an anisotropic behavior occurs, presumably resulting from a contribution of $Ni$ thin film spin orbit coupling. The main magnetoresistive signal, however, is attributed to the higher resistance given by the random orientation of crystal vortex polarizations and chiralities at zero field, which enhance the resistance due to the higher density of scattering and spin mixing events. The lower resistance observed near the saturation magnetization in the longitudinal and transverse configurations (in-plane) as well as the vortex core orientation with the external field (out-of-plane) is provided by the onset of a low resistive path caused by vortex core orientation or in-plane saturation by the out-of-plane external magnetic field. Similar behavior was recently observed for systems with vortex array \cite{japclusters}. The magnetoresitive measurements evidently show long-range spin polarized transport throughout the system. In Fig. \ref{fig6} we present results for spin polarization calculated from MR measurements performed in same configurations as in Fig. \ref{fig5} and in different temperatures. The highest polarization was achieved upon out-of-plane magnetic field sweeping, corroborating our observations of crystal vortice with random chirality and core polarization at low magnetic field, as highest spin mixing resistive path, and successive low resistive path obtained by alignment of vortex polarization and chirality with the field. The observed linear decrease of polarization in function of temperature, is in good agreement with the spin diffusion length behavior in metallic non-magnetic spacer in multilayer magnetoresitive devices \cite{sdlferromagnetic}, corroborating our model.


In summary, ferromagnetic resonance measurements were used to assess the anisotropy behavior of as-fabricated antidot systems, which is responsible for crystal vortex formation, as it was recently predicted in the literature. The crystal vortex configuration was furthermore confirmed by ferromagnetic resonance measurements with the onset of magnon states signature at higher magnetic fields. Additional experimental evidence of the vortex crystal created by the insertion of the antidots in the nickel thin film was assessed by local magnetoresistance measurements. The anisotropic magnetoresistance expected for thin magnetic films is non-zero around zero magnetic field and the main isotropic peaks are related to the generation of the vortex crystal. The long-range spin polarization is highest at low temperatures and decreases linearly while with increasing temperature, as it is expected for metallic ferromagnetic thin films.      

\begin{acknowledgments}
The authors thank CNPq, CAPES and FAPEMIG (Brazilian agencies) for
financial support. They also would like to thank the LABNANO/CBPF for technical support during electron microscopy/nanolithography work.
\end{acknowledgments}

\end{document}